\documentclass[a4paper]{llncs}
\usepackage{amsmath}
\usepackage{amssymb}
\usepackage{graphicx}
\usepackage{algorithm2e}
\usepackage{breqn}
\usepackage{multirow}
\usepackage{capt-of}
\usepackage[T1]{fontenc}
\usepackage{letltxmacro}

\usepackage{url}
\urldef{\mailsa}\path|{alfred.hofmann, ursula.barth, ingrid.haas, frank.holzwarth,|
\urldef{\mailsb}\path|anna.kramer, leonie.kunz, christine.reiss, nicole.sator,|
\urldef{\mailsc}\path|erika.siebert-cole, peter.strasser, lncs}@springer.com|    
\newcommand{\keywords}[1]{\par\addvspace\baselineskip
\noindent\keywordname\enspace\ignorespaces#1}

  \graphicspath{{Figures/}}
   \DeclareGraphicsExtensions{.pdf,.jpg,.png}

\begin{document}

\mainmatter  % start of an individual contribution

\title{Recovering Architectural Variability of a Family of Product Variants\vspace{-0.80em}}

\titlerunning{Recovering Architectural Variability Model of a Familly of Product Variants} 


\author{Anas Shatnawi\inst{1} \and Abdelhak Seriai\inst{1}
 \and Houari Sahraoui\inst{2}}

\institute{UMR CNRS 5506, LIRMM, University of Montpellier II, Montpellier, France\\
\email{{shatnawi, seriai}@lirmm.fr}
%\texttt{http://users/\homedir iekeland/web/welcome.html}
\and
DIRO, University of Montreal, Montreal, Canada\\
\email{sahraoui@iro.umontreal.ca\vspace{-1.7em}}
}

\maketitle

\begin{abstract}
A Software Product Line (SPL) aims at applying a pre-planned systematic reuse of large-grained software artifacts to increase the software productivity and reduce the development cost. The idea of SPL is to analyze the business domain of a family of products to identify the common and the variable parts between the products. However, it is common for companies to develop, in an ad-hoc manner (e.g. clone and own), a set of products that share common functionalities and differ in terms of others. Thus, many recent research contributions are proposed to re-engineer existing product variants to a SPL. Nevertheless, these contributions are mostly focused on managing the variability at the requirement level. Very few contributions address the variability at the architectural level despite its major importance. Starting from this observation, we propose, in this paper, an approach to reverse engineer the architecture of a set of product variants. Our goal is to identify the variability and dependencies among architectural-element variants at the architectural level. Our work relies on Formal Concept Analysis (FCA) to analyze the variability. To validate the proposed approach, we experimented on two families of open-source product variants; Mobile Media and Health Watcher. The results show that our approach is able to identify the architectural variability and the dependencies.
\end{abstract}
 \vspace{-2.2em}
 \vspace{-0.9em}
 
\section{Introduction}
\label{introduction}
\vspace{-0.5em}
%A Software Product Line (SPL) is \textquotedblleft a set of software-intensive systems that share a common, managed set of features satisfying the specific needs of a particular market segment or mission and that are developed from a common set of core assets in a prescribed way\textquotedblright \cite{1_clements2002}. 
A Software Product Line (SPL) aims at applying a pre-planned systematic reuse of large-grained software artifacts (e.g. components) to increase the software productivity and reduce the development cost \cite{1_clements2002,3_pohl2005software,4_tan2012quality}. The main idea behind SPL is to analyze the business domain of a family of products in order to identify the common and the variable parts between these products \cite{1_clements2002,3_pohl2005software}.
In SPL, the variability is realized at different levels of abstraction (e.g. requirement and design). At the requirement level, it is originated starting from the differences in users' wishes, and does not carry any technical sense \cite{3_pohl2005software} (e.g. the user needs \textit{camera} and \textit{WIFI} features in the phone). At the design level, the variability starts to have more details related to technical senses to form the product architectures. These technical senses are described via Software Product Line Architecture (SPLA). Such technical senses are related to which components compose the product (e.g. \textit{video recorder}, and \textit{photo capture} components), how these components interact through their interfaces (e.g. \textit{video recorder} provides a \textit{video stream} interface to \textit{media store}), and what topology forms the architectural configuration (i.e. how components are composited and linked) \cite{3_pohl2005software}.

Developing a SPL from scratch is a highly costly task since this means the development of the domain software artifacts \cite{1_clements2002}. In addition, it is common for companies to develop a set of software product variants that share common functionalities and differ in terms of other ones. These products are usually developed in an ad-hoc manner (e.g. clone and own) by adding or/and removing some functionalities to an existing software product to meet the requirement of a new need \cite{10_rubin2012locating}. Nevertheless, when the number of product variants grows, managing the reuse and maintenance processes becomes a severe problem \cite{10_rubin2012locating}. As a consequence, it is necessary to identify and manage variability between product variants as a SPL. The goal is to reduce the cost of SPL development by first starting it from existing products and then being able to manage the reuse and maintenance tasks in product variants using a SPL. Thus, many research contributions have been proposed to re-engineer existing product variants into a SPL \cite{23_she2011reverse,25_acher2011reverse}. Nevertheless, existing works are mostly focused on recovering the variability in terms of features defined at the requirement level. Despite the major importance of the SPLA, there is only two works aiming at recovering the variability at the architectural level \cite{koschke2009extending,kang2005feature}. These approaches are not fully-automated and rely on the domain knowledge which is not always available. Also, they do not identify dependencies among the architectural elements. To address this limitation, we propose in this paper an approach to automatically recover the architecture of a set of software product variants by capturing the variability at the architectural level and the dependencies between the architectural elements. We rely on Formal Concept Analysis (FCA) to analyze the variability. In order to validate the proposed approach, we experimented on two families of open-source product variants; Mobile Media and Health Watcher. The evaluation shows that our approach is able to identify the architectural variability and the dependencies as well.

 %SPLA is recognized as an important building unit of SPL engineering \cite{1_clements2002,2_galster2011handling,3_pohl2005software,4_tan2012quality}. SPLA aims to capture the variability of a family of product architectures. It describes how architectural elements can be configured to form a concrete architecture of a family member \cite{2_galster2011handling,4_tan2012quality}.
 
The rest of this paper is organized as follows. Section \ref{Background} presents the background needed to understand our proposal. Then, in Section \ref{Process}, we present the recovery process of SPLA. Section \ref{IdentifyingArchitectureVariability} presents the identification of architecture variability. Then, Section \ref{IdentifyingArchitectureDependencies} presents the identification of dependencies among architectural-element variants. Experimental evaluation of our approach is discussed in section \ref{Experimentation}. Then, the related work is discussed in Section \ref{RelatedWork}. Finally, concluding remarks and future directions are presented in section \ref{Conclusion}.
\vspace{-1.2em}
\section{Background}
\vspace{-0.6em}
\label{Background}

%\subsection{Variability in Software Product Line}
%\vspace{-0.5em}
%\label{VariabilityinSPL}

%The main theme in SPL is the variability. It is related to the susceptibility and flexibility to change. The variability in SPL is realized at different levels of abstraction during the development life cycle (e.g. requirement and design). At the requirement level, it is originated starting from the differences in users' wishes, and does not carry any technical sense \cite{3_pohl2005software} (e.g. the user needs camera and WIFI features in the phone). At the design level, the variability starts to have more details related to technical senses to form the product architectures. These technical senses describe how the products are built with regard to the point of view of software architects \cite{3_pohl2005software}. Such technical senses are related to which components compose the product (e.g. video recorder,  photo capture, and media store components), how these components interact through their interfaces (e.g. video recorder provides a video stream interface to media store), and what topology forms the architectural configuration (i.e. how components are composited and linked) \cite{3_pohl2005software}. All these technical senses are described via SPLA.
%\vspace{-1.5em}
\subsection{Component-Based Architecture Recovery from Single Software: ROMANTIC Approach}
\begin{figure}
\vspace{-2.0em}
\centering
\includegraphics[width=7cm]{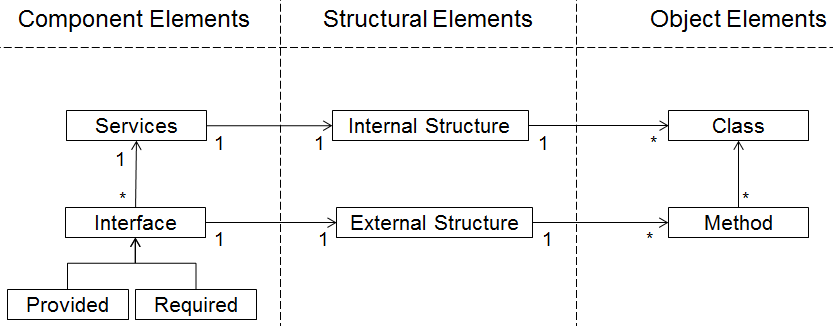}
    \caption{Object-to-component mapping model}
\label{fig:mappingModel1}
\vspace{-2.9em}
\end{figure}

\vspace{-0.7em}
In our previous work \cite{9_kebir2012quality,8_chardigny2008extraction}, \textit{ROMANTIC}\footnote{\textit{ROMANTIC}: Re-engineering of Object-oriented systeMs by Architecture extractioN and migraTIon to Component based ones.} approach has been proposed to automatically recover a component-based architecture from the source code of an existing object-oriented software. Components are obtained by partitioning classes of the software. Thus each class is assigned to a unique subset forming a component. \textit{ROMANTIC} is based on two main models. The first concerns the object-to-component mapping model which allows to link object-oriented concepts (e.g. package, class) to component-based ones (e.g. component, interface). A component consists of two parts; internal and external structures. The internal structure is implemented by a set of classes that have direct links only to classes that belong to the component itself. The external structure is implemented by the set of classes that have direct links to other components' classes. Classes that form the external structure of a component define the component interface.
Fig. \ref{fig:mappingModel1} shows the object-to-component mapping model. The second main model proposed in \textit{ROMANTIC} is used to evaluate the quality of recovered architectures and their architectural-element. For example, the quality-model of recovered components is based on three characteristics; composability, autonomy and specificity. These refer respectively to the ability of the component to be composed without any modification, to the possibility to reuse the component in an autonomous way, and to the fact that the component implements a limited number of closed functionalities. Based on these models, \textit{ROMANTIC} defines a fitness function applied in a hierarchical clustering algorithm \cite{9_kebir2012quality,8_chardigny2008extraction} as well as in search-based algorithms \cite{28_chardigny2008search} to partition the object-oriented classes into groups, where each group represents a component. In this paper, \textit{ROMANTIC} is used to recover the architecture of a single object oriented software product. 

\vspace{-1.0em}
\subsection{Formal Concept Analysis}
\label{sec:FCA}
\vspace{-0.5em}
Formal Concept Analysis (FCA) is a mathematical data analysis technique developed based on lattice theory \cite{6_ganter1996formal}. It allows the analysis of the relationships between a set of objects described by a set of attributes. In this context, maximal groups of objects sharing the same attributes are called formal concepts. These are extracted and then hierarchically organized into a graph called a concept lattice. Each formal concept consists of two parts. The first allows the representation of the objects covered by the concepts called the extent of the concept. The second allows the representation of the set of attributes shared by the objects belonging to the extent. This is called the intent of the concept. Concepts can be linked through sub-concept and super-concept relationship \cite{6_ganter1996formal} where the lattice defines a partially ordered structure. A concept $A$ is a sub-concept of the super-concept $B$, if the extent of the concept $B$ includes the extent of the concept $A$ and the intent of the concept $A$ includes the intent of the concept $B$. 
\begin{minipage}{\textwidth}
  \begin{minipage}[b]{0.30\textwidth}
    \centering
    \begin{tabular}{|c|c|c|c|c|c|c|c|} \hline
             & \rotatebox{270} {Natural} & \rotatebox{270}{Artificial} & \rotatebox{270}{Stagnant}	& \rotatebox{270}{Running} & \rotatebox{270}{Inland} &	\rotatebox{270}{Maritime} & \rotatebox{270}{Constant}\\ \hline
            River& X& & & X & X & & X \\ \hline
            Sea& X & & X&  & & X & X \\ \hline
            Reservoir& & X& X & & X & & X \\ \hline
            Channel& & & & X & X & & X \\ \hline
            Lake& X& & X &  & X & & X  \\ \hline
            \end{tabular}
      \captionof{table}{Formal context}
      \label{formalContexExample}
    \end{minipage}
    \hfill
    \begin{minipage}[b]{0.55\textwidth}
        \centering
        \includegraphics[width=0.85\textwidth]{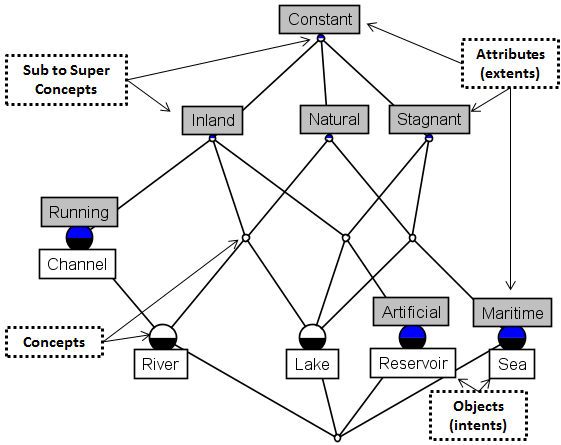}
        \captionof{figure}{Lattice of formal context in Table \ref{formalContexExample}}
        \label{FCA_Water1}
      \end{minipage}
\vspace{0.6em}
  \end{minipage}
The input of FCA is called a formal context. A formal context is defined as a triple $K = (O, A, R)$ where $O$ refers to a set of objects, $A$ refers to a set of attributes and $R$ is a binary relation between objects and attributes. This binary relation indicates to a set of attributes that are held by each object (i.e. $R\subseteq O X A$).  Table \ref{formalContexExample} shows an example of a formal context for a set of bodies of waters and their attributes. An $X$ refers to that an object holds an attribute.

As stated before, a formal concept consists of extent $E$ and intent $I$, with $E$ a subset of objects $O$ $(E\subseteq O)$  and $I$ a subset of attributes $A$ $(I\subseteq A)$. A pair of extent and intent $(E, I)$ is considered a formal concept, if and only, if $E$ consists of only objects that shared all attributes in $I$ and $I$ consists of only attributes that are shared by all objects in $E$. The pair ("river, lake", "inland, natural, constant") is an example of a formal concept of the formal context in Table \ref{formalContexExample}. Fig. \ref{FCA_Water1} shows the concept lattice of the formal context presented in Table \ref{formalContexExample}.

\vspace{-1.2em}
\section{Process of Recovering Architectural Variability}
\label{Process}
\vspace{-0.6em}
The goal of our approach is at recovering the architectural variability of a set of product variants by statically analyzing their object-oriented source code. This is obtained by identifying variability among architectures respectively recovered from each single product. We rely on \textit{ROMANTIC} approach to extract the architecture of a single product. This constitutes the first step of the recovery process. Architecture variability is related to architectural-elements variability, i.e. component, connector and configuration variability. In our approach, we focus only on component and configuration variability\footnote{Most of architectural description languages do not consider connector as a first class concept.}. Fig. \ref{fig:archVarExam1} shows an example of architecture variability based on component and configuration variability. In this example, there are three product variants, where each one diverges in the set of component constituting its architecture as well as the links between the components. Component variability refers to the existence of many variants of one component. \textit{CD Reader} and \textit{CD Reader / Writer} represent variants of one component. We identify component variants based on the identification of components providing similar functionalities. This is the role of the second step of the recovery process. Configuration variability is represented in terms of presence/absence of components on the one hand (e.g. \textit{Purchase Reminder}), and presence/absence of component-to-component links on the other hand (e.g. the link between \textit{MP3 Decoder / Encoder} and \textit{CD Reader / Writer}). We identify configuration variability based on both the identification of core (e.g. \textit{Sound Source}) and optional components (e.g. \textit{Purchase Reminder}) and links between these components. In addition, we capture the dependencies and constraints among components. This includes, for example, require constraints between optional components. We rely on FCA to identify these dependencies. These are mined in the fourth step of the recovery process. Fig. \ref{fig:miningProc1} shows these steps.
\vspace{-1.0em}
\section{Identifying the Architecture Variability} 
\label{IdentifyingArchitectureVariability}
\vspace{-0.5em}
The architecture variability is mainly materialized either through the existence of variants of the same architectural element (i.e. component variants) or through the configuration variability. In this section, we show how component variants and configuration variability are identified.
\vspace{-1.0em}
\subsection{Identifying Component Variants}
\begin{figure}
\vspace{-1.0em}
\centering
\parbox{4.05cm}{
\includegraphics[width=4.15cm]{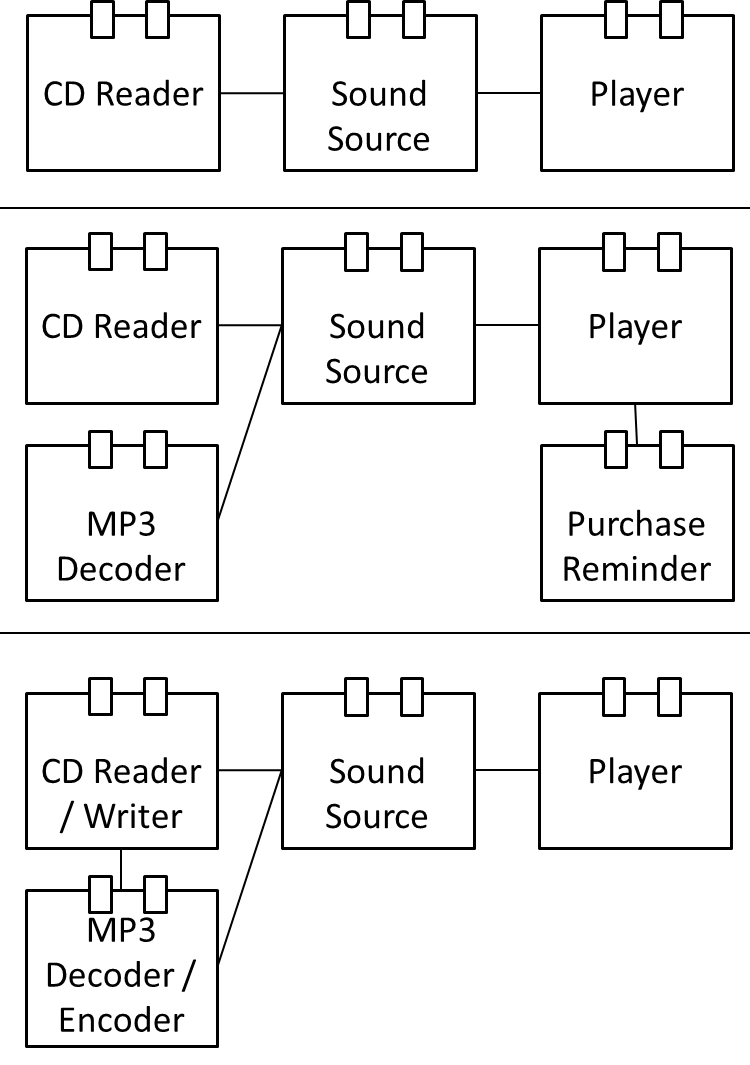}
\caption{An example of architecture variability}
\label{fig:archVarExam1}}
\qquad
\begin{minipage}{6.4cm}
\includegraphics[width=6.4cm]{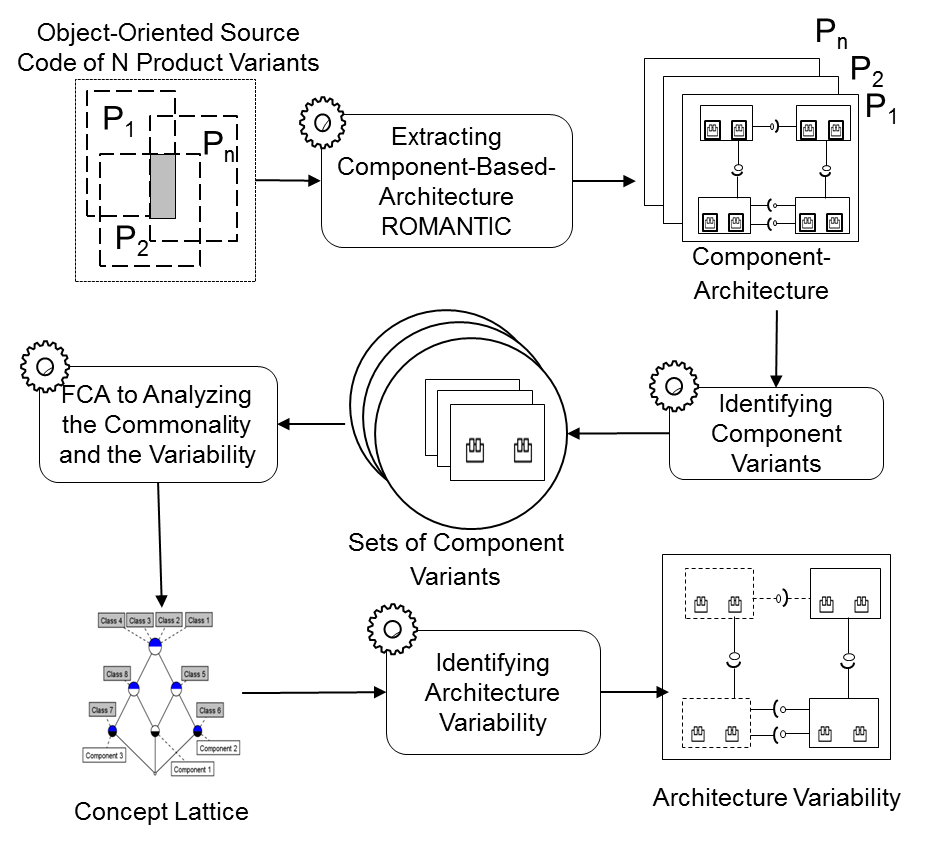}
\caption{The process of architectural variability recovery}
\label{fig:miningProc1}
\end{minipage}
\vspace{-2.0em}
\end{figure}

\vspace{-0.5em}
The selection of a component to be used in an architecture is based on its provided and required services. The provided services define the role of the component. However, other components may provide the same, or similar, core services. Each may also provide other specific services in addition to the core ones. Considering these components, either as completely different or as the same, does not allow the variability related to components to be captured. Thus, we consider them as component variants. We define component variants as a set of components providing the same core services and differ concerning few secondary ones. In Fig. \ref{fig:archVarExam1},  \textit{MP3 Decoder} and \textit{MP3 Decoder / Encoder} are component variants. 

We identify component variants based on their similarity. Similar components are those sharing the majority of their classes and differing in relation to some others. Components are identified as similar based on the strength of similarity links between their implementing classes. For this purpose, we use cosine similarity metric \cite{11_han2006data} where each component is considered as a text document composed of the names of its classes. We use a hierarchical clustering algorithm \cite{11_han2006data} to gather similar components into clusters. It starts by considering components as initial leaf nodes in a binary tree. Next, the two most similar nodes are grouped into a new one that forms their parent. This grouping process is repeated until all nodes are grouped into a binary tree. All nodes in this tree are considered as candidates to be selected as groups of similar components. To identify the best nodes, we use a depth first search algorithm. Starting from the tree root to find the cut-off points, we compare the similarity of the current node with its children. If the current node has a similarity value exceeding the average similarity value of its children, then the cut-off point is in the current node. Otherwise, the algorithm continues through its children. The results of this algorithm are clusters where each one is composed of a set of similar components that represent variants of one component.
\vspace{-1.5em}
\subsection{Identifying Configuration Variants}
\vspace{-0.5em}
The architectural configuration is defined based on the list of components composing the architecture, as well as the topology of the links existing between these components. Thus the configuration variability is related to these two aspects; the lists of core (mandatory) and optional components and the list of core and optional links between the selected components.
\vspace{-1.5em}
\subsubsection{Identification of component variability:}
To identify mandatory and optional components, we use Formal Concept Analysis (FCA) to analyze architecture configurations. We present each software architecture as an object and each member component as an attribute in the formal context. In the concept lattice, common attributes are grouped into the root while the variable ones are hierarchically distributed among the non-root concepts.
\begin{figure}
\vspace{-1.0em}
\centering
\includegraphics[scale=0.35]{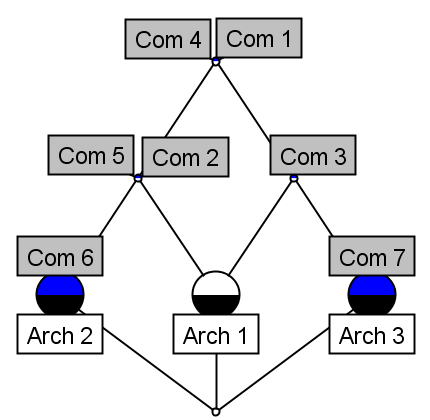}
\caption{A lattice example of similar configurations}
\label{fig:lattArchExam}
\vspace{-2.3em}
\end{figure}

Fig. \ref{fig:lattArchExam} shows an example of a lattice for three similar architecture configurations. The common components (the core ones) are grouped together at the root concept of the lattice (the top). In Fig. \ref{fig:lattArchExam} \textit{Com1} and \textit{Com4} are the core components present in the three architectures. By contrast, optional components are represented in all lattice concepts except the root. e.g., according to the lattice of Fig. \ref{fig:lattArchExam},\textit{Com2} and \textit{Com5} present in \textit{Arch1} and \textit{Arch2} but not in \textit{Arch3}.
\vspace{-1.5em}
\subsubsection{Identification of component-link variability:}

A component-link is defined as a connection between two components where each connection is the abstraction of a group of  method invocation, access attribute or inheritance links between classes composing these components. In the context of configuration variability, a component may be linked with different sets of components. A component may have links with a set of components in one product, and it may have other links with a different set of components in another product. Thus the component-link variability is related to the component variability. This means that the identification of the link variability is based on the identified component variability. For instance, the existence of a link \textit{A-B} is related to the selection of a component \textit{A} and a component \textit{B} in the architecture. Thus considering a core link (mandatory link) is based on the occurrence of the linked components, but not on the occurrence in the architecture of products. According to that, a core link is defined as a link occurring in the architecture configuration as well the linked components are selected. To identify the component-link variability, we proceed as follows. For each architectural component, we collect the set of components that are connected to it in each product. The intersection of the sets extracted from all the products determines all core links for the given component. The other links are optional ones.
\vspace{-1.5em}
\section{Identifying Architecture Dependencies}
\label{IdentifyingArchitectureDependencies}
\vspace{-0.7em}
The identification of component and component-link variability is not enough to define a valid architectural configuration. This also depends on the set of dependencies (i.e. constraints) that may exist between all the elements of the architecture. For instance, components providing antagonism functionalities have an exclude relationship. Furthermore, a component may need other components to perform its services. Dependencies can be of five kinds:  alternative, OR, AND, require, and exclude dependencies. To identify them we rely on the same concept lattice generated in the previous section. 

In the lattice, each node groups a set of components representing the intent and a set of architectural configurations representing the extent. The configurations are represented by paths starting from their concepts to the lattice concept root. The idea is that each object is generated starting from its node up going to the top. This is based on sub-concept to super-concept relationships (c.f. Section \ref{sec:FCA}). This process generates a path for each object. A path contains an ordered list of nodes based on their hierarchical distribution; i.e. sub-concept to super-concept relationships). According to that, we propose extracting the dependencies between each pair of nodes as follows:
%\vspace{-0.5em}
\begin{itemize}
	\item \textbf{Required dependency.} This constraint refers to the obligation selection of a component to select another one; i.e. component \textit{B} is required to select component \textit{A}. Based on the generated lattice, we analyze all its nodes by identifying parent-to-child relation (i.e. top to down). Thus node \textit{A} requires node \textit{B} if node \textit{B} appears before node \textit{A} in the lattice, i.e., node \textit{A} is a sub-concept of the super-concept corresponding to node \textit{B}. In other words, to reach node \textit{A} in the lattice, it is necessary to traverse node \textit{B}. For example, if we consider lattice of the Fig. \ref{fig:lattArchExam}, \textit{Com6} requires \textit{Com2} and \textit{Com5} since \textit{Com2} and \textit{Com5} are traversed before \textit{Com6} in all paths including \textit{Com6} and linking root node to object nodes.

	\item \textbf{Exclude and alternative dependencies.} Exclude dependency refers to the antagonistic relationship; i.e. components \textit{A} and \textit{B} cannot occur in the same architecture. This relation is extracted by checking all paths linking root to all leaf nodes in the lattice. A node is excluded with respect to another node if they never appear together in any of the existing paths; i.e. there is no sub-concept to super-concept relationship between them. This means that there exists no object exists containing both nodes. For example, if we consider lattice of Fig. \ref{fig:lattArchExam}, \textit{Com6} and \textit{Com7} are exclusives since they never appear together in any of the lattice paths.
	
	Alternative dependency generalizes the exclude one by exclusively selecting only one component from a set of components. It can be identified based on the exclude dependencies. Indeed, a set of nodes in the lattice having each an exclude constraint with all other nodes forms an alternative situation.
	
	\item \textbf{AND dependency.}  This is the bidirectional version of the REQUIRE constraint; i.e. component \textit{A} requires component \textit{B} and vice versa. More generally, the selection of one component among a set of components requires the selection of all the other components. According to the built lattice, this relation is found when a group of components is grouped in the same concept node in the lattice; i.e. the whole node should be selected and not only a part of its components. For example if we consider lattice of the Fig. \ref{fig:lattArchExam}, \textit{Com2} and \textit{Com5} are concerned with an AND dependency.
	
	\item \textbf{OR dependency.} When some components are concerned by an OR dependency, this means that at least one of them should be selected; i.e. the configuration may contain any combination of the components. Thus, in the case of absence of other constraints any pair of components is concerned by an OR dependency.  Thus pairs concerned by required, exclude, alternative, or AND dependencies are ignored as well as those concerned by transitive require constraints; e.g. \textit{Com6} and \textit{Com7} are ignored since they are exclusives. Algorithm \ref{algo:ORMining} shows the procedure of identifying groups of OR dependency. 
\end{itemize}

\begin{algorithm}

\SetAlgoLined
\caption{Identifying OR-Groups}
\label{algo:ORMining}
\SetKwFunction{match}{match}
 \KwIn{all pairs (ap), require dependencies (rd), exclude dependencies (ed) and alternative dependencies (ad)}
 \KwOut{sets of nodes having OR dependencies (orGroups)}
 OrDep = ap.exclusionPairs(rd, ed, ad);\\
 OrDep = orDep.removeTransitiveRequire(rd);\\
 ORPairsSharingNode = orDep.getPairsSharingNode();\\
 
 \For{each pairs p in ORPairsSharingNode}
  {
  \uIf{otherNodes.getDependency() == require}{orDep.removePair(childNode);}
      \uElseIf{otherNodes.getDependency()= exclude || alternative}{orDep.removeAllPairs(p);}
  }
  orGroups = orDep.getpairssharingOrDep();\\
\KwRet{orGroups}\\
\end{algorithm}
\vspace{-1.6em}
\section{Experimentation and Results}
\label{Experimentation}
\vspace{-0.6em}
Our experimentation aims at showing how the proposed approach is applied to identify the architectural variability and validating the obtained results. To this end, we applied it on two case studies. We select two sets of product variants. These sets are Mobile Media\footnote{Available at: http://ptolemy.cs.iastate.edu/design-study/\#mobilemedia.} (MM) and Health Watcher\footnote{Available at: http://ptolemy.cs.iastate.edu/design-study/\#healthwatcher.} (HW). We select these products because they were used in many research papers aiming at addressing the problem of migrating product variants into a SPL. Our study considers 8 variants of MM and 10 variants of HW. MM variants manipulate music, video and photo on mobile phones. They are developed starting from the core implementation of MM. Then, the other features are added incrementally for each variant. HW variants are web-based applications that aim at managing health records and customer complaints. The size of each variant of MM and HW, in terms of classes, is shown in Table \ref{systemSizes1}. We utilize \textit{ROMANTIC} approach \cite{9_kebir2012quality} to extract architectural components from each variant independently. Then, the components derived from all variants are the input of the clustering algorithm to identify component variants. Next, we identify the architecture configurations of the products. These are used as a formal context to extract a concept lattice. Then, we extract the core (mandatory) and optional components as well as the dependencies among optional-component. 

In order to evaluate the resulted architecture variability, we study the following research questions:
%\vspace{-0.5em}
\begin{itemize}
%\item \textbf{RQ1: Do the identified component variants represent the same architectural element?}
%The aim of this research question is at proving our hypothesis concerning component variants by checking how much components sharing the most of their classes represent the same architectural element. 
\item \textbf{RQ1: Are the identified dependencies correct?}
This research question goals at measuring the correctness of the identified component dependencies.
\item \textbf{RQ2: What is the precision of the recovered architectural variability?}
This research question focuses on measuring the precision of the resulting architecture variability. This is done by comparing it with a pre-existed architecture variability model.
\end{itemize}
\begin{table}[ht]
\vspace{-2.0em}
\centering
\caption{Size of MM variants and HW ones}
\label{systemSizes1}
\begin{tabular}{|c|c|c|c|c|c|c|c|c|c|c|c|} \hline
 Name&	1&	2&	3&	4&	5&	6&	7&	8&	9&	10&	Avg.\\ \hline
MM&	25&	34&	36&	36&	41&	50&	60&	64&	X&	X&	43.25 \\ \hline
HW&	115&	120&	132&	134	& 136&	140&	144&	148&	160&	167&	136.9 \\ \hline
\end{tabular}
\vspace{-2.0em}
\end{table}

\begin{figure}
\vspace{-1.5em}
\centering
\includegraphics[scale=0.50]{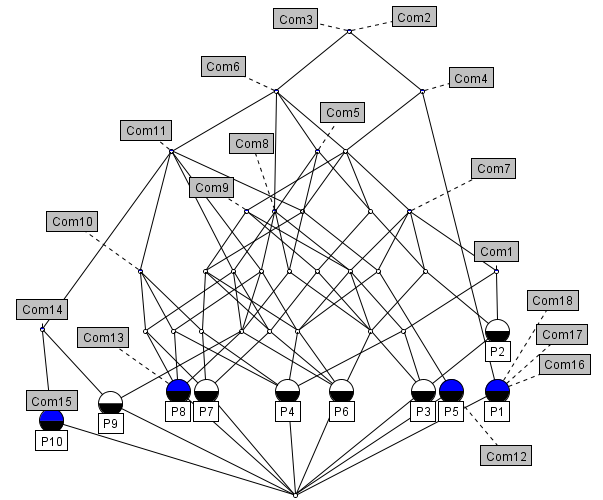}
    \caption{The concept lattice of HW architecture configurations}
\label{fig:HWLattice}
\vspace{-2.0em}
\end{figure}
\vspace{-0.5em}
\subsection{Results}
\vspace{-0.2em}
 Table 3 shows the results of component extraction from each variant independently, in terms of the number of components, for each variant of MM and HW. The results show that classes related to the same functionality are grouped into the same component. The difference in the numbers of the identified components in each variant has resulted from the fact that each variant has a different set of user's requirements. On average, a variant contains 6.25 and 7.7 main functionalities respectively for MM and HW.

\begin{table}
\vspace{-2.0em}
\parbox{.45\linewidth}{
\centering
\label{components2}
\caption{Comp. extraction results}
\begin{tabular}{|c|c|c|c|c|c|c|c|c|c|c|c|c|} \hline
 Name&	1&	2&	3&	4&	5&	6&	7&	8&	9&	10&	Avg. & Total\\ \hline
MM&	3&	5&	5&	5&	7&	7&	9&	9&	X&	X&	6.25&	50\\ \hline
HW&	6&	7&	9&	10&	7&	9&	8&	8&	7&	6&	7.7&	77\\ \hline
\end{tabular}
}
\hfill
\parbox{.45\linewidth}{
\centering
\label{comVar}
\caption{Comp. variants identification}
\begin{tabular}{|c|c|c|c|c|} \hline
 Name&	NOCV&	ANVC&	MXCV&	MNCV\\ \hline
MM&	14&	3.57&	8&	1\\ \hline
HW&	18&	4.72&	10&	1\\ \hline
\end{tabular}
}
\vspace{-2.0em}
\end{table}

Table 4 summarizes the results of component variants in terms of the number of components having variants (NOCV), the average number of variants of a component (ANVC), the maximum number of component variants (MXCV) and the minimum number of component variants (MNCS). The results show that there are many sets of components sharing the most of their classes. Each set of components mostly provides the same functionality. Thus, they represent variants of the same architectural component. Table \ref{comvar} presents an instance of 6 component variants identified from HW, where \textit{X} means that the corresponding class is a member in the variant. By analyzing these variants, it is clear that these components represent the same architectural component. In addition to that, we noticed that there are some component variants having the same set of classes in multiple product variants.

\begin{table}[ht]
\vspace{-0.5em}
\centering
\caption{Instance of 6 component variants}
\resizebox{0.75\textwidth}{!}{
\label{comvar}
\begin{tabular}{|c|c|c|c|c|c|c|} \hline
 Class Name & Variant 1&	Variant 2&	Variant 3 &	Variant 4&	Variant 5&	Variant 6\\ \hline
BufferedReader&	X&	X & X & X& X& X\\ \hline
ComplaintRepositoryArray & X&	X & X& X& X & X\\ \hline
ConcreteIterator&	X&	X & X& X& X& X\\ \hline
DiseaseRecord&	X & & & & &\\ \hline
IIteratorRMITargetAdapter&	X&	X & X& X& X& X\\ \hline
IteratorRMITargetAdapter&	X&	X & X& X& X& X\\ \hline
DiseaseType& & X& & & &\\ \hline
InputStreamReader& X & X& X& X& X& X\\ \hline
Employee& & X& & X& &\\ \hline
InvalidDateException & & & X& X & X & X\\ \hline
IteratorDsk& X&	X& X& X& X& X\\ \hline
PrintWriter& X&	X& X& & X & X\\ \hline
ObjectNotValidException & & & X& & X& X\\ \hline
RemoteException& X& X& & X& &\\ \hline
PrintStream& & & X& & X& X\\ \hline
RepositoryException& X& X & & & &\\ \hline
Statement&	X &	X& X& X & X & X\\ \hline
Throwable&	X &	X& & X& &\\ \hline
HWServlet&	 &	& & & X&\\ \hline
Connection& & & & & X& X\\ \hline
\end{tabular}
}
\vspace{-2.0em}
\end{table}

The architecture configurations are identified based on the above results. Table 6 shows the configuration of MM variants, where \textit{X} means that the component is a part of the product variants. The results show that the products are similar in their architectural configurations and differ considering other ones. The reason behind the similarity and the difference is the fact that these products are common in some of their user's requirements and variable in some others. These architecture configurations are used as a formal context to extract the concept lattice. We use the Concept Explorer\footnote{Presentation of the Concept Explorer tool is available in \cite{14_Yevtushenko2000}.} tool to generate the concept lattice. Due to limited space, we only give the concept lattice of HW (c.f. Fig. \ref{fig:HWLattice}). In Table 7, the numbers of core (mandatory) and optional components are given for MM and HW. The results show that there are some components that represent the core architecture, while some others represent delta (optional) components. %The core architecture provides functionalities shared by all product variants.

\begin{table}
\vspace{-2.0em}
\parbox{.45\linewidth}{
\centering
\label{ArchCong}
\caption{Arch. configuration for all MM variants}
\resizebox{0.9\textwidth}{!}{\begin{minipage}{\textwidth}
\begin{tabular}{|c|c|c|c|c|c|c|c|c|c|c|c|c|c|c|} \hline
Variant No.&	\rotatebox{270}{Com1}	& \rotatebox{270}{Com2}&	\rotatebox{270}{Com3}&	\rotatebox{270}{Com4}&	\rotatebox{270}{Com5}&	\rotatebox{270}{Com6}&	\rotatebox{270}{Com7}&	\rotatebox{270}{Com8}&	\rotatebox{270}{Com9}&	\rotatebox{270}{Com10}&	\rotatebox{270}{Com11}&	\rotatebox{270}{Com12}&	\rotatebox{270}{Com13}&	\rotatebox{270}{Com14}\\ \hline
1& X& & X& & X& & & & & & & & &  \\ \hline
2& X& & X& & X& & & X& X& & & & &  \\ \hline
3& X& & X& & X& & & X& X& & & & &  \\ \hline
4& & & X& & X& & & X& X& & X& & &  \\ \hline
5& X& & X& & X& & & X& X& & X& X& &  \\ \hline
6& & & X& & X& & X& X& X& & X& X& &  \\ \hline
7& & X& & X& X& & X& X& X& X& X& X& &   \\ \hline
8& & X& & X& X& X& X& X& & X& & & X& X  \\ \hline
\end{tabular}
\end{minipage} }
}
\hfill
\parbox{.45\linewidth}{
\centering
\label{FinalResults}
     \caption{Mandatory and optional components}
\begin{tabular}{|c|c|c|} \hline
     Product Name&	MM&	HW \\\hline
    Mandatory&	1&	2 \\\hline 
    Optional&	13&	16 \\\hline
     \end{tabular}
}
\vspace{-2.0em}
\end{table}

The results of the identification of optional-component dependencies are given in Table 8 (\textit{Com 5} is excluded since it is a mandatory component). For conciseness, the detailed dependencies among components are only shown for MM only. The dependencies are represented between all pairs of components in MM (where R= Require, E= Exclude, O= OR, RB = Required By, TR = Transitive Require, TRB = Transitive Require By, and A = AND). Table 9 shows a summary of MM and HW dependencies between all pairs of components. This includes the number of direct require constrains (NRC), the number of exclude ones (NE), the number of AND groups (NOA), and the number of OR groups (NO). Alternative constrains is represented as exclude ones. The results show that there are dependencies among components that help the architect to avoid creating invalid configuration. For instance, a design decision of AND components indicates that these components depend on each other, thus, they should be selected all together.

\begin{table}
\parbox{.45\linewidth}{
\centering
\vspace{-2.0em}
\label{constriant}
\caption{Component dependencies}
\resizebox{0.75\textwidth}{!}{\begin{minipage}{\textwidth}
\begin{tabular}{|c|c|c|c|c|c|c|c|c|c|c|c|c|c|c|} \hline
  & C1& C2& C3& C4& C6& C7& C8& C9 & C10& C11& C12& C13& C14\\ \hline
 Com1& X& & R& & E& E& & & & O& E& E& E\\ \hline
 Com2& & X& E& A& RB& R& TR& & A& & & RB& RB\\ \hline
 Com3& RB& E& X& E& E& & O& & E& & & E& E\\ \hline
 Com4& & A& E& X& RB& R& TR& & A& & & RB& RB\\ \hline
 Com6& E& R& E& R& X& TR& TR& E& R& E& E& A& A\\ \hline
 Com7& E& RB& & RB& TRB& X& R& O& RB& & & TRB& TRB\\ \hline
 Com8& & TRB& O& TRB& TRB& RB& X& RB& TRB& TRB& TRB& TRB& TRB\\ \hline
 Com9& & & & & E& O& R& X& & RB& TRB& E& E\\ \hline
 Com10& & A& E& A& RB& R& TR& & X& & & RB& RB\\ \hline
 Com11& O& & & & E& & TR& R& & X& RB& E& E\\ \hline
 Com12& E& & & & E& & TR& TR& & R& X& E& E\\ \hline
 Com13& E& R& E& R& A& TR& TR& E& R& E& E& X& A\\ \hline
 Com14& E& R& E& R& A& TR& TR& E& R& E& E& A& X\\ \hline
 \end{tabular}
\end{minipage} }
}
\hfill
\parbox{.35\linewidth}{
\centering
\label{sumOfDep}
     \caption{Summarization of MM and HW dependencies}
\begin{tabular}{|c|c|c|c|c|} \hline
         Name& NDR& NE& NA& NO\\ \hline
        MM& 17& 20& 6& 3\\ \hline
        HW& 18& 62& 3& 11\\ \hline
        \end{tabular}
}
\vspace{-2.0em}
\end{table}

\begin{figure}[h]
\vspace{-0.5em}
  \begin{center}
    \includegraphics[scale=0.33]{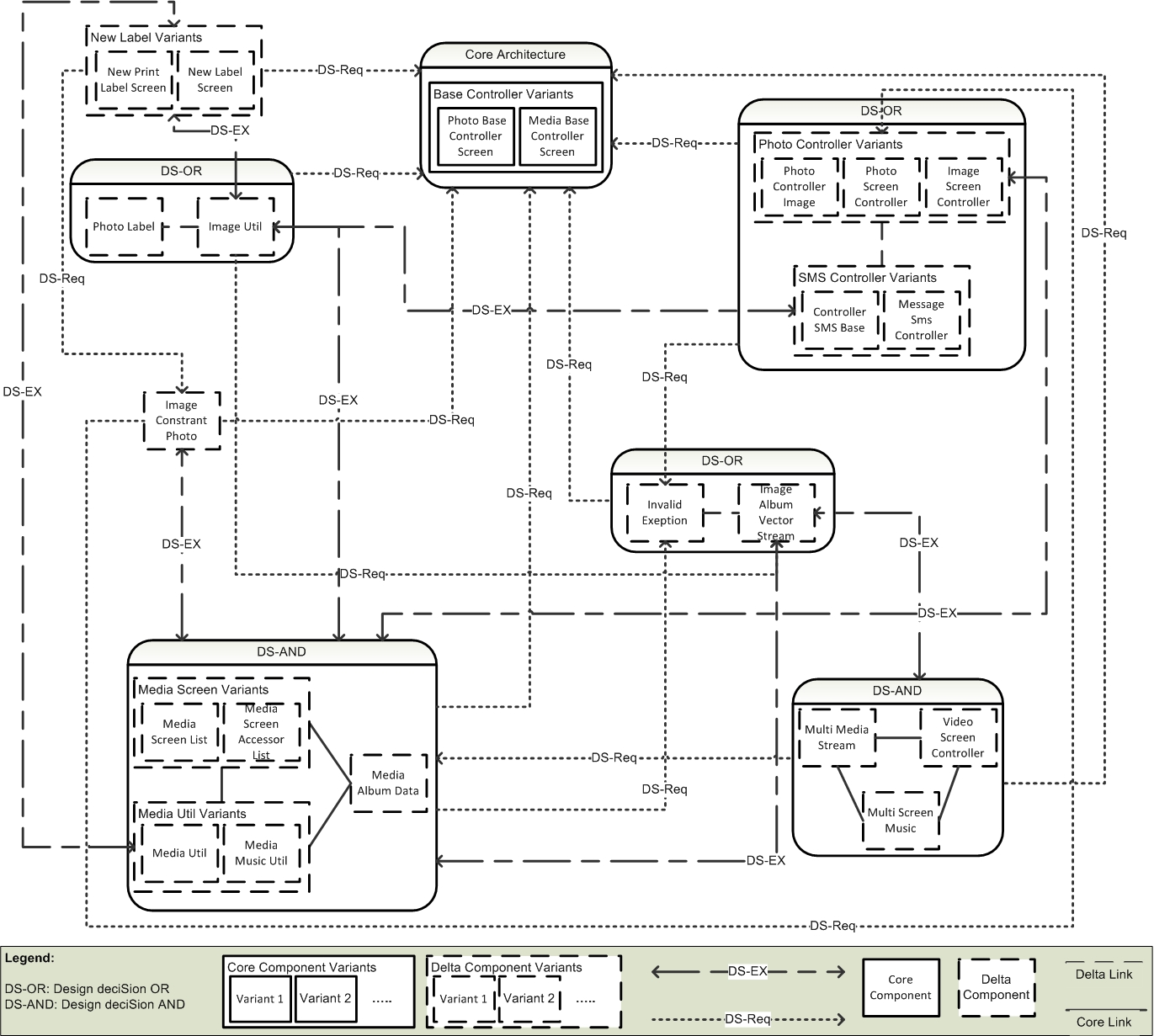}
    \caption{Architectural variability model for MM}
        \label{MMmodel}
  \end{center}
\vspace{-3.2em}
\end{figure}

To the best our knowledge, there is no architecture description language supporting all kinds of the identified variability. The existing languages are mainly focused on modeling component variants, links and interfaces, while they do not support dependencies among components such as  AND-group, OR-group, and require. Thus, on the first hand, we use some notation presented in \cite{Hendrickson:2007} to represent the concept of component variants and links variability. On the other hand, we propose some notation inspired from feature modeling languages to model the dependencies among components. For the purpose of understandability, we document the resulting components by assigning a name based on the most frequent tokens in their classes' names. Figure \ref{MMmodel} shows the architectural variability model identified for MM variants, where the large boxes denote to design decisions (constraints). For instance, core architecture refers to components that should be selected to create any concrete product architecture. In MM, there is one core components manipulating the base controller of the product. This component has two variants. A group of \textit{Multi Media Stream}, \textit{Video Screen Controller}, and \textit{Multi Screen Music} components represents an AND design decision.

%The notations are as follows: solid boxes refer to core architecture elements, dashed boxes denote to optional elements, large rectangles contain component variants, big boxes attached with a title refer to design decisions of OR or AND group of components, .%

%\subsubsection{RQ1: Do the identified component variants represent the same architectural element?}
%To validate how much component variants represent the same architectural element, we compare their implementation. To this end, we randomly select a set of component variants. Then, we manually check attribute names, method names, and method bodies of classes having the same name of these components. Present the result....
\vspace{-1.5em}
\subsubsection{RQ1: Are the identified dependencies correct?}
The identification of component dependencies is based on the occurrence of components. e.g., if two components never selected to be included in a concrete product architecture, we consider that they hold an exclude relation. However, this method could provide correct or incorrect dependencies. To evaluate the accuracy of this method, we manually validate the identified dependencies. This is based on the functionalities provided by the components. For instance, we check if the component functionality requires the functionality of the required component and so on. The results show that 79\% of the required dependencies are correct. As an example of a correct relation is that \textit{SMS Controller} requires \textit{Invalid Exception} as it performs an input/output operations. On the other hand, it seems that \textit{Image Util} does not require \textit{Image Album Vector Stream}. Also, 63\% of the exclude constrains are correct. For AND and OR dependencies, we find that 88\% of AND groups are correct, while 42\% of OR groups are correct. Thus, the precision of identifying dependencies is 68\% in average.
\vspace{-1.5em}
\subsubsection{RQ2: What is the precision of the recovered architectural variability?}
\begin{figure}[h]
\vspace{-1.7em}
  \begin{center}
    \includegraphics[scale=0.46]{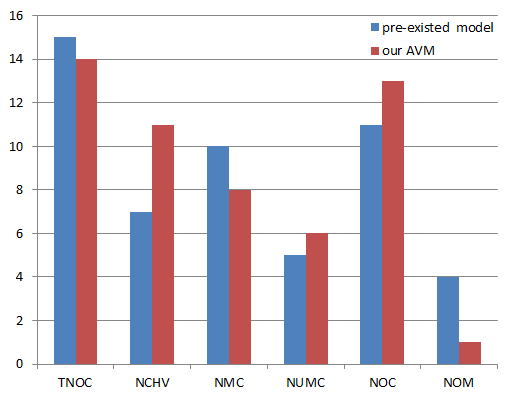}
    \caption{The results of the MM validation}
        \label{vallidation}
  \end{center}
  \vspace{-4.5em}
\end{figure}

In our case studies, MM is the only case study that has an available architecture model containing some variability information. In \cite{27_figueiredo2008evolving}, the authors presented the aspect oriented architecture for MM variants. This contains information about which products had added components, as well as in which product a component implementation was changed (i.e. component variants). We manually compare both models to validate the resulting model. Fig. \ref{vallidation} shows the comparison results in terms of the total number of components in the architecture model (TNOC), the number of components having variants (NCHV), the number of mapped components in the other model (NC), the number of unmapped components in the other model (NUMC), the number of optional components (NOC) and the number of mandatory ones (NOM). The results show that there are some variation between the results of our approach and the pre-existed model. The reason behind this variation is the idea of compositional components. For instance, our approach identifies only one core component compared to 4 core components in the other model. Our approach grouped all classes related to the controller components together in one core components. On the other hand, the other model divided the controller component into \textit{Abstract Controller}, \textit{Album Data}, \textit{Media Controller}, and \textit{Photo View Controller} components. In addition, the component related to handling exceptions is not mentioned in the pre-existed model at all.

\section {Related Work} 
\vspace{-0.9em}
\label{RelatedWork}
In this section, we discuss the contributions that have been proposed in two research directions; recovering the software architecture of a set of product variants and variability management. 

In \cite{koschke2009extending}, an approach aiming at recovering SPLA was presented. It identifies component variants based on the detection of cloned code among the products. However, the limitation of this approach is that it is a semi-automated, while our approach is fully automated. Also, it does not identify dependencies among the components. In \cite{kang2005feature}, the authors presented an approach to reconstruct Home Service Robots (HSR) products into a SPL. Although this approach identifies some architectural variability, but it has some limitation compared to our approach. For instance, it is specialized on the domain of HSR as the authors classified, at earlier step, the architectural units based on three categories related to HSR. These categories guide the identification process. In addition, the use of feature modeling language (hierarchical trees) to realize the identified variability is not efficient as it is not able to represent the configuration of architectures. Domain knowledge plays the main role to identify the architecture of each single product and the dependencies among components. In some cases, domain knowledge is not always available. The authors in \cite{25_acher2011reverse} proposed an approach to reverse engineering architectural feature model. This is based on the software architect's knowledge, the architecture dependencies, and the feature model that is extracted based on a reverse engineering approach presented in \cite{23_she2011reverse}. The idea, in \cite{25_acher2011reverse}, is to take the software architect's variability point of view in the extracted feature model (i.e. still at the requirement level); this is why it is named architecture feature model. However, the major limitations of this approach are firstly that the software architect is not available in most cases of legacy software, and secondly that the architecture dependencies are generally missing as well. In \cite{23_she2011reverse}, the authors proposed an approach to extract the feature model. The input of the extraction process is feature names, feature descriptions and dependencies among features. Based on this information, they recover ontological constraints (e.g. feature groups) and cross tree constrains. A strong assumption behind this approach is that feature names, feature descriptions, and dependencies among features are available. In \cite{26_ryssel2011extraction}, the authors use FCA to generate a feature model. The input of their approach is a set of feature configurations. However, the extraction of the feature model elements is based on NP-hard problems (e.g. set cover to identify or groups). Furthermore, architecture variability is not taken into account in this approach. In \cite{21_duszynski2011analyzing}, an approach was presented to visually analyze the distribution of variability and commonality among the source code of product variants. The analysis includes multi-level of abstractions (e.g. line of code, method, class, etc.). This aims to facilitate the interpretation of variability distribution, to support identifying reusable entities. In \cite{shatnawi:IRI:2013}, the authors presented an approach to extract reusable software components from a set of similar software products. This is based on identifying similarity between components identified independently from each software. This approach can be related only to the first step of our approach.

%In \cite{20_klatt2012respecting}, the authors proposed an approach to reverse engineer the variability of product variants. This approach uses the information about component architecture and the requirements to improve the variability management. The authors rely on the abstract syntax tree model to identify the variation points among the variants. The identified variation points are then merged, based on the architectural information. The limitation of this approach is that, in addition to considering the availability of the requirement and the architecture information, the comparison of the abstract syntax trees of all variants does not scale for large systems. The authors in \cite{22_wu2011recovering} proposed a semi-automatic approach to analyze the variability of a set of similar legacy products. They depend on object-oriented models and their implementation to identify the variation points. First, the variability is identified based on method and class levels, using a clone detection technique. Then, they classify the identified variability elements.

\vspace{-1.5em}
\section{Conclusion}
\label{Conclusion}
\vspace{-0.9em}
%Integrating variability via SPLA is a vital task for the software architect. 
In SPLA, the variability is mainly represented in terms of components and configurations. In the case of migrating product variants to a SPL, identifying the architecture variability among the product variants is necessary to facilitate the software architect's tasks. Thus, in this paper, we proposed an approach to recover the architecture variability of a set of product variants. The recovered variability includes mandatory and optional components, the dependencies among components (e.g. require, etc.), the variability of component-links, and component variants. We rely on FCA to analyze the variability. Then, we propose two heuristics. The former is to identify the architecture variability. The latter is to identify the architecture dependencies. The proposed approach is validated through two sets of product variants derived from Mobile Media and Health Watcher. The results show that our approach is able to identify the architectural variability and the dependencies as well.

There are three aspects to be considered regarding the hypothesis of our approach. Firstly, we identify component variants based on the similarity between the name of classes composing the components, i.e., classes that have the same name should have the same implementation. While in some situations, components may have very similar set of classes, but they are completely unrelated. Secondly, dependencies among components are identified based on component occurrences in the product architectures. Thus, the identified dependencies maybe correct or incorrect. Finally, the input of our approach is the components independently identified form each product variants using \textit{ROMANTIC} approach. Thus. the accuracy of the obtained variability depends on the accuracy of \textit{ROMANTIC} approach. 

Our future research will focus on migrating product variants into component based software product line, the mapping between the requirements' variability (i.e. features) and the architectures' variability, and mapping between components' variability and component-links' variability.

%\bibliographystyle{splncs}
%\bibliography{./Biblio/BiblioCSMR}

\vspace{-1.0em}

\begin{thebibliography}{4}

 \bibitem{1_clements2002} Clements, P., Northrop, L.: Software product lines: practices and patterns. Addison-Wesley Reading (2002)
 
 \bibitem{3_pohl2005software} Pohl, K. and B{\"o}ckle, G. and Van Der Linden, F.: Software product line engineering. Springer Berlin Heidelberg (2005)
 
 \bibitem{4_tan2012quality} Tan, L. and Lin, Y. and Ye, H.: Quality-oriented software product line architecture design. Journal of Software Engineering \& Applications. 5(7), 472--476 (2012)
 
 \bibitem{10_rubin2012locating} Rubin, J., Chechik, M.: Locating distinguishing features using diff sets. In: IEEE/ACM 27th Inter. Conf. on ASE, pp. 242--245. (2012)
 
 \bibitem{23_she2011reverse} She, S., Lotufo, R., Berger, T., Wasowski, A., Czarnecki, K.: Reverse engineering
 feature models. In: Proc. of 33rd ICSE, pp. 461--470. (2011)
  
  \bibitem{25_acher2011reverse} Acher, M., Cleve, A., Collet, P., Merle, P., Duchien, L., Lahire, P.: Reverse engineering
  architectural feature models. In: Software Architecture. LNCS, vol. 6903, pp. 220--235. Springer, Heidelberg (2011)  
  
  \bibitem{koschke2009extending} Koschke, R., Frenzel, P., Breu, A.P., Angstmann, K.: Extending the reflexion
  method for consolidating software variants into product lines. Software Quality
  Journal. 17(4), 331--366 (2009)
  
  \bibitem{kang2005feature} Kang, K.C., Kim, M., Lee, J., Kim, B.: Feature-oriented re-engineering of legacy
  systems into product line assets - a case study. In: Software Product Lines. LNCS, vol. 3714, pp. 45--56. Springer, Heidelberg (2005)  
 
 \bibitem{9_kebir2012quality} Kebir, S., Seriai, A.D., Chardigny, S., Chaoui, A.: Quality-centric approach for software component identification from object-oriented code. In: Proc. of WICSA/ECSA, pp. 181--190. (2012)     
   
 \bibitem{8_chardigny2008extraction} Chardigny, S., Seriai, A., Oussalah, M., Tamzalit, D.: Extraction of componentbased
   architecture from object-oriented systems. In: Proc. of 7th WICSA, pp. 285--288. (2008)
 
 \bibitem{28_chardigny2008search} Chardigny, S., Seriai, A.D., Oussalah, M., Tamzalit, D.: Search-based extraction
   of component-based architecture from object-oriented systems. In: 2nd ECSA. LNCS, vol. 5292, pp. 322--325. Springer, Heidelberg (2008)
     
   \bibitem{6_ganter1996formal} Ganter, B., Wille, R.: Formal concept analysis. WISSENSCHAFTLICHE
   ZEITSCHRIFT-TECHNISCHEN UNIVERSITAT DRESDEN. 47, 8--13 (1996)
     
   \bibitem{11_han2006data} Han, J., Kamber, M., Pei, J.: Data mining: concepts and techniques. Morgan kaufmann (2006)
   
   \bibitem{14_Yevtushenko2000} Yevtushenko, A.S.: System of data analysis "concept explorer". (In Russian) Proc. of the 7th National Conf. on Artificial Intelligence (KII), vol. 79, pp. 127--134. (2000)
   
    \bibitem{Hendrickson:2007} Hendrickson, S.A., van der Hoek, A.: Modeling product line architectures through change sets and relationships. In: Proc. of the 29th ICSE, pp. 189--198. (2007)
       
    \bibitem{27_figueiredo2008evolving} Figueiredo, E. and Cacho, N. and Sant'Anna, C. and Monteiro, M. and Kulesza, U. and Garcia, A. and Soares, S. and Ferrari, F. and Khan, S., et al.: Evolving software product lines
    with aspects. In: Proc. of 30th ICSE, pp. 261--270. (2008)
              
    \bibitem{26_ryssel2011extraction} Ryssel, U., Ploennigs, J., Kabitzsch, K.: Extraction of feature models from formal
    contexts. In: Proc. of 15th SPLC, pp. 1--4. (2011)
     
     \bibitem{21_duszynski2011analyzing} Duszynski, S., Knodel, J., Becker, M.: Analyzing the source code of multiple
     software variants for reuse potential. In: Proc. of WCRE, pp. 303--307. (2011)      
           
    \bibitem{shatnawi:IRI:2013} Shatnawi, A., Seriai, A.D.: Mining reusable software components from objectoriented
       source code of a set of similar software. In: IEEE 14th Inter. Conf. on
       Information Reuse and Integration (IRI), pp. 193--200. (2013)
      
      
 
 \end{thebibliography}
%\bibitem{jour} Smith, T.F., Waterman, M.S.: Identification of Common Molecular
%Subsequences. J. Mol. Biol. 147, 195--197 (1981)
%\bibitem{lncschap} May, P., Ehrlich, H.C., Steinke, T.: ZIB Structure Prediction Pipeline:
%ùComposing a Complex Biological Workflow through Web Services. In: Nagel,
%ùW.E., Walter, W.V., Lehner, W. (eds.) Euro-Par 2006. LNCS, vol. 4128,
%pp. 1148--1158. Springer, Heidelberg (2006)
%\bibitem{book} Foster, I., Kesselman, C.: The Grid: Blueprint for a New Computing
%Infrastructure. Morgan Kaufmann, San Francisco (1999)
%\bibitem{proceeding1} Czajkowski, K., Fitzgerald, S., Foster, I., Kesselman, C.: Grid
%Information Services for Distributed Resource Sharing. In: 10th IEEE
%International Symposium on High Performance Distributed Computing, pp.
%181--184. IEEE Press, New York (2001)
%\bibitem{proceeding2} Foster, I., Kesselman, C., Nick, J., Tuecke, S.: The Physiology of the
%Grid: an Open Grid Services Architecture for Distributed Systems
%Integration. Technical report, Global Grid Forum (2002)
%\bibitem{url} National Center for Biotechnology Information, \url{http://www.ncbi.nlm.nih.gov}
%\end{thebibliography\}

\end{document}